\documentclass[aps,prb,showpacs,twocolumn]{revtex4}
\usepackage{graphicx}

\begin{document}
\title{Phase diagrams of a classical two-dimensional
       Heisenberg antiferromagnet with single-ion anisotropy}
\date{8 September, 2004}
\author{R.~Leidl}
\author{W.~Selke}
\affiliation{Institut f\"ur Theoretische Physik, Technische Hochschule,
             52056 Aachen, Germany}
 
\begin{abstract}
  A classical variant of the two-dimensional anisotropic Heisenberg model
  reproducing inelastic neutron scattering experiments on
  La$_5$Ca$_9$Cu$_{24}$O$_{41}$
  [M. Matsuda \textit{et al.}, Phys.\ Rev.\ B \textbf{68}, 060406(R) (2003)]
  is analysed using mostly Monte Carlo techniques. Phase diagrams
  with external fields parallel and perpendicular to the easy axis
  of the anisotropic interactions are determined, including antiferromagnetic
  and spin-flop phases. Mobile spinless defects, or holes, are found
  to form stripes which bunch, debunch and break up at a phase transition.
  A parallel field can lead to a spin-flop phase.
\end{abstract}


\maketitle

\section{Introduction}
The compounds (La,Ca)$_{14}$Cu$_{24}$O$_{41}$ have been studied experimentally
rather extensively in recent years.\cite{ca,m1,miz,m2,am,ka,windt,rk,m4}
They display interesting low-di\-mensional magnetic properties
arising from Cu$_2$O$_3$ two-leg ladders and CuO$_2$ chains.
In addition, for La$_{14-x}$Ca$_x$Cu$_{24}$O$_{41}$ there is an intriguing
interplay between spin and charge ordering due to hole doping when $x>8$.

In particular, La$_5$Ca$_9$Cu$_{24}$O$_{41}$ exhibits, at low temperatures
and small fields, antiferromagnetic long range order associated
with the CuO$_2$ chains which are oriented along the $c$ axis.
The Cu$^{2+}$ ions in those chains carry a spin-1/2.
The spins are ordered ferromagnetically in the chains,
while the interchain coupling in the $ac$--planes is antiferromagnetic.
The magnetic interactions between the $ac$--planes are believed
to be very weak. The couplings in the $ac$--planes are anisotropic
with an easy axis along the $b$ axis, i.e.\ the magnet
has an Ising-like character.\cite{am,ka,rk,m4}
Holes may originate from the La and Ca ions, transforming Cu ions
into spinless quantities, with a hole content of about 10 percent.\cite{nu}

Experiments on La$_5$Ca$_9$Cu$_{24}$O$_{41}$ include thermodynamic
measurements on the specific heat, magnetization,
and susceptibility \cite{am,rk} as well as electron spin resonance \cite{ka}
and neutron scattering.\cite{miz,rk} Motivated by the experimental findings,
different models have been proposed and studied.
Firstly, a two-dimensional Ising model has been introduced,\cite{spbk,hs}
where the spins correspond to the Cu$^{2+}$ ions, and mobile spinless defects
mimic the holes.
In addition to the ferromagnetic intrachain and antiferromagnetic interchain
couplings between neighboring spins, a rather strong and antiferromagnetic
exchange between next-nearest chain spins separated by a hole is assumed.
The model has been shown to describe, at low temperatures, antiferromagnetic
domains separated by quite straight defect lines which break up
at a phase transition where also the long range magnetic order gets destroyed.
The stripe instability is caused by an effectively attractive interaction
between the defects mediated by the antiferromagnetic interchain couplings.

Even more recently, a two-dimensional anisotropic Heisenberg model has been
shown by Matsuda \textit{et al.}\ to reproduce the measured spin-wave
dispersions which supposedly result from the collective spin excitations
of the Cu$^{2+}$ ions in the $ac$--planes.\cite{m4}
Our subsequent Monte Carlo simulations on the classical variant of the model
seem to indicate, however, that some of its thermodynamic properties
tend to deviate from experimental findings.\cite{ls} In particular,
in an external field along the $b$ axis, at low temperatures,
the field dependence of the susceptibility of the anisotropic Heisenberg model
disagrees with the measured behavior. The disagreement can probably
not be resolved by invoking quantum effects. In particular,
the critical temperature of the model seems to be significantly
lower than the measured one, as can be seen by comparing results on related
quantum and classical models.\cite{ls,cuc,se} Indeed, in this paper
we will show that the qualitative properties of the phase diagram of the model
are not affected qualitatively by quantum effects. 

In any event, the classical variant of the model
of Matsuda \textit{et al.}\ deserves to be studied in more detail.
Apart from the previous  qualitative comparison \cite{ls} with experimental
data, the classical model is of genuine theoretical interest as well.
Perhaps most interestingly, the phase diagram in the
(temperature,field)--plane is worth studying in the context
of two-dimensional anisotropic Heisenberg models. Due to the anisotropy,
there is a spin-flop phase when applying a sufficiently high external field
along the $b$ axis. In two dimensions, that phase has interesting properties
as it is believed to be of Kosterlitz-Thouless type
with spin correlations decaying algebraically with distance.
Furthermore, the boundary line of the antiferromagnetic phase
as well as the transition between the antiferromagnetic and
the spin-flop phase has been discussed controversially for anisotropic
nearest-neighbor Heisenberg models. In three and higher dimensions,
there is a bicritical point where the antiferromagnetic, the spin-flop,
and the paramagnetic  phases meet.\cite{fn} As it is well known,
such a point could occur in two dimensions only at zero temperature
(Mermin-Wagner theorem). Actually, different scenarios have been
proposed.\cite{bl} However, simulations and other analyses
on classical models lead to controversial results.\cite{bl,cp}
Fairly recently, a $S=1/2$ quantum version  has been simulated suggesting
a topology of the phase diagram with a tricritical
and a critical end point.\cite{sch} Our work will deal with this aspect.
Note that quasi two-dimensional anisotropic Heisenberg antiferromagnets
exhibiting a spin-flop phase have attracted much experimental attention
already some years ago.\cite{jon,rau,stei} Well-known examples
are Rb$_2$MnCl$_4$ and K$_2$MnF$_4$. The approach of our study
may be also useful for the correct theoretical analysis of models
describing these materials.   

Our main emphasis will be on a classical variant of the model by Matsuda
\textit{et al.}\ obtained from the spin-wave dispersions
of La$_5$Ca$_9$Cu$_{24}$O$_{41}$.
In addition, in an attempt to include possible effects due to the holes,
we shall extend the model by allowing for mobile defects following
the previous considerations.\cite{spbk} In fact, the defects again form
stripes which are destabilized at a phase transition. Bunching and debunching
of the stripes are novel features. Effects of an external field
along the $b$ axis will also be considered.

The layout of the article is as follows: In the next section, the model
will be introduced. Its phase diagrams, without defects
and applying external fields parallel and perpendicular to the easy axis
of the magnetic interactions, will be presented in Sec.\ \ref{sec_phdiag}.
The possible influence of defects on thermal properties will be discussed
in section \ref{sec_defects}, followed by a summary.

\section{The model, simulations, and quantities of interest}\label{sec_model}
\begin{figure}[h]
  \includegraphics{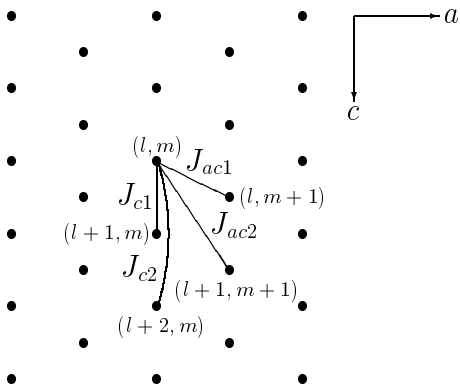}
  \caption{Sketch of the the magnetic interactions for the two-dimensional
   anisotropic Heisenberg model of Matsuda \textit{et al.}\cite{m4}
   The dots denote the sites of the Cu$^{2+}$ ions in the $ac$--plane of
   La$_5$Ca$_9$Cu$_{24}$O$_{41}$.}
  \label{fig_Matmodel}
\end{figure}

Following Matsuda \textit{et al.},\cite{m4} the magnetic properties
of La$_5$Ca$_9$Cu$_{24}$O$_{41}$ depend on the $\text{Cu}^{2+}$
ions located in the $ac$--planes, having
a centered rectangular geometry as depicted in Fig.\ \ref{fig_Matmodel}.
Based on their spin-wave analysis, the spins ($S=1/2$)
of the ions couple along the CuO$_2$ chains, i.e.\ along the $c$ axis,
through nearest neighbor, $J_{c1}$, and next-nearest neighbor, $J_{c2}$,
exchange constants, with $J_{c1}=-0.2$ meV being antiferromagnetic
and $J_{c2}=0.18$ meV being ferromagnetic.
The ferromagnetic ordering in the chains is due to the strong
antiferromagnetic interchain couplings, see also Fig.\ \ref{fig_Matmodel}:
$J_{ac1}=-0.681$ meV refers to the two nearest neighbors
in the adjacent chain, and  $J_{ac2}=0.5J_{ac1}=-0.3405$ meV denotes
the couplings to the two next-nearest neighbors.  

Importantly, there is an uniaxial exchange anisotropy favoring alignment
of the spins along the $b$ axis. Its contribution to the different couplings
cannot be determined in the spin-wave analysis, and only its integral effect
on the gap in the dispersion is quantified.\cite{m4} When going over
to a classical description with spins of fixed length, say, one, the total
anisotropy may be mimicked by a single-ion term. Such a term would be,
of course, merely a trivial constant and unphysical for a quantum system
with $S$= 1/2. The single-ion term, avoiding ambiguities in distributing
the anisotropy among the different couplings, is quite reasonable
in the classical variant. It leads to about the same transition temperature
without field as various exchange anisotropies of the same overall
magnitude.\cite{ls} In addition, the entire phase diagram seems
to be not affected by the type of anisotropy.
The anisotropy has been estimated from the gap energy to be
$D=-0.211$ meV.\cite{m4} Thence the Hamiltonian can be written as \cite{m4,ls}
\begin{widetext}
  \begin{eqnarray}
    {\cal H} &=&
     - J_{c1} \sum\limits_{(l,m)} \mathbf{S}_{l,m} \mathbf{S}_{l+1,m}
     - J_{c2} \sum\limits_{(l,m)} \mathbf{S}_{l,m} \mathbf{S}_{l+2,m}
     - J_{ac1}\sum\limits_{(l,m)} \mathbf{S}_{l,m}
        (\mathbf{S}_{l,m+1}+\mathbf{S}_{l+1,m+1})\nonumber\\
    \label{Ham}
    & &
     {}-J_{ac2} \sum\limits_{(l,m)} \mathbf{S}_{l,m} (\mathbf{S}_{l-1,m+1}
      + \mathbf{S}_{l+2,m+1}) 
      + D\sum\limits_{(l,m)}( S_{l,m}^z)^2
      - H_{\alpha} \sum\limits_{(l,m)} S_{l,m}^{\alpha},
  \end{eqnarray}
\end{widetext}
where $\mathbf{S}_{l,m}=(S_{l,m}^x,S_{l,m}^y,S_{l.m}^z)$ denotes the spin
at the $l$--th site in the $m$--th chain; $\alpha$ refers to differently
oriented magnetic fields, with $\alpha=z$ for the field parallel
to the easy axis, and $\alpha=x$ or $y$
for a field perpendicular to it, i.e.\ along the $a$ or $c$ axis.

Extending this Hamiltonian of Matsuda \textit{et al.}, we introduce defects
$S_{l,m}=0$, induced by the holes originating
from the La and Ca ions.\cite{spbk} We neglect direct interactions
between the defects, and there are no couplings between a defect and a spin.
Next-nearest neighbor spins in the same chain, $S_{l,m}$ and $S_{l\pm2,m}$,
with a defect in between them are coupled by the exchange constant $J_0$
(replacing $J_{c2}$), which we presume to be antiferromagnetic
and rather strong as before, i.e.\ of several meV.\cite{spbk}
Specifically, we choose $J_0=-6.25$ meV.
Thence, the Hamiltonian (\ref{Ham}) is augmented by the term  
\begin{equation}
  \label{Ham_defects}
  {\cal H}_d = -J_0 \sum\limits_{(l,m)} S_{l,m} S_{l+2,m} (1- n_{l+1, m}),
\end{equation}
with $n_{l,m}=0,1$ being the occupation variable of a spin at site $(l,m)$.
The defects are allowed to hop to a neighboring site in a chain,
transforming the spin at that site into a defect and leaving a new spin
with arbitrary orientation at its initial site. The probability
of such a process is determined by the Boltzmann factor of the change
in the magnetic energy, Eqs.\ (\ref{Ham}) and (\ref{Ham_defects}),
associated with the hop.\cite{spbk} It is easily incorporated
in Monte Carlo simulations.

As before,\cite{spbk} we assume that defects are separated along the chain
by at least one spin. The number of defects in each chain will be taken
to be 10 percent of the number of sites in that chain.
The defect concentration is then close to that
in La$_5$Ca$_9$Cu$_{24}$O$_{41}$.\cite{nu,rk}

We shall study the model, with and without defects, using,
apart from ground state calculations, standard Monte Carlo techniques
with local elementary processes changing at randomly chosen sites
the spin orientation and moving defects to neighboring sites.
In the simulations, we consider lattices with the same number $L$ of chains
and of sites per chain, i.e.\ with a total of $L^2$ sites.
Full periodic boundary conditions are employed. To study finite-size effects
allowing extrapolations to the thermodynamic limit, $L\longrightarrow\infty$,
we consider typically sizes ranging from $L$=10 to $L$=200. Each run consists
of at least $10^6$ and up to $5\times10^6$ Monte Carlo steps per site.
To obtain averages and error bars, we take into account the results
of up to 10 realizations using different random numbers.

We compute both quantities of direct experimental interest as well
as other quantities which enable us to determine critical properties
and the phase transition lines. In particular, we recorded
the specific heat $C$, both from the fluctuations and from the temperature
derivative of the energy per site $E$. In the absence of defects,
various magnetizations were computed. Especially, we recorded

\noindent
(i) the $\alpha$--component of the magnetization to study the response
to a field in the $\alpha$ direction, with $\alpha=x,y,z$,
\begin{equation}
  <M^{\alpha}> = <\sum\limits_{(l,m)} S_{l,m}^{\alpha}>/L^2; 
\end{equation}

\noindent
(ii) the $z$--component of the absolute value of the staggered magnetization
$M_s^z$ and the square of the staggered magnetization
to describe the order in the antiferromagnetic phase, 
\begin{equation}
  \label{Mzstagg}
  <|M_s^z|> =
   <|\sum\limits_{(\mathrm{all}\,l,\,m\,\mathrm{even})}S_{l,m}^z
    - \sum\limits_{(\mathrm{all}\,l,\,m\,\mathrm{odd})}S_{l,m}^z|>/L^2,
\end{equation}
(summing separately over sites in even and odd chains),
and similarly for $<(M_s^z)^2>$; and

\noindent
(iii) the square of the transverse sublattice magnetization to describe
the Kosterlitz-Thouless character of the spin-flop phase when applying
a field along the easy axis,
\begin{widetext}
  \begin{equation}
    \label{Mxy}
    <M_{xy}^2> = <\sum\limits_{\alpha=x,y}\Big[
       \Big(\sum\limits_{(\mathrm{all}\,l,\,m\,\mathrm{even})}
        S_{l,m}^{\alpha}\Big)^2
     + \Big(\sum\limits_{(\mathrm{all}\,l,\,m\,\mathrm{odd})}
        S_{l,m}^{\alpha}\Big)^2 \Big]> \big/ (L^4/2).
  \end{equation}
\end{widetext}
In addition, we recorded the magnetic susceptibilities $\chi^{\alpha}$,
which may be computed from the fluctuations or field derivatives
of the corresponding magnetizations, $<M^{\alpha}>$,
and the (finite lattice) staggered susceptibility $\chi_s^z$ defined by
\begin{equation}
  \chi_s^z = L^2\,\big(<(M_s^z)^2> - <|M_s^z|>^2\big)/2.
\end{equation}
To identify the type of transition from the antiferromagnetic
to the paramagnetic and the spin-flop phases, the fourth-order,
size dependent cumulant of the staggered magnetization,
the Binder cumulant \cite{bin} $U_L$, is rather useful: 
\begin{equation}
  \label{BindCum}
  U_L = 1- <(M_s^z)^4>/\big(3 <(M_s^z)^2>^2\big),
\end{equation}
where $<(M_s^z)^4>$ is defined in analogy to $<(M_s^z)^2>$.

In the presence of defects we studied, apart from the specific heat,
magnetization and susceptibility, also microscopic quantities describing
the topology and stability of the defect stripes,\cite{spbk,hs}
including the average minimal distance between defects in adjacent chains
and the density of defect pairs.
Results will be discussed in Sec.\ \ref{sec_defects}.

\section{Phase diagrams in the absence of defects}\label{sec_phdiag}
\begin{figure}[h]
  \includegraphics{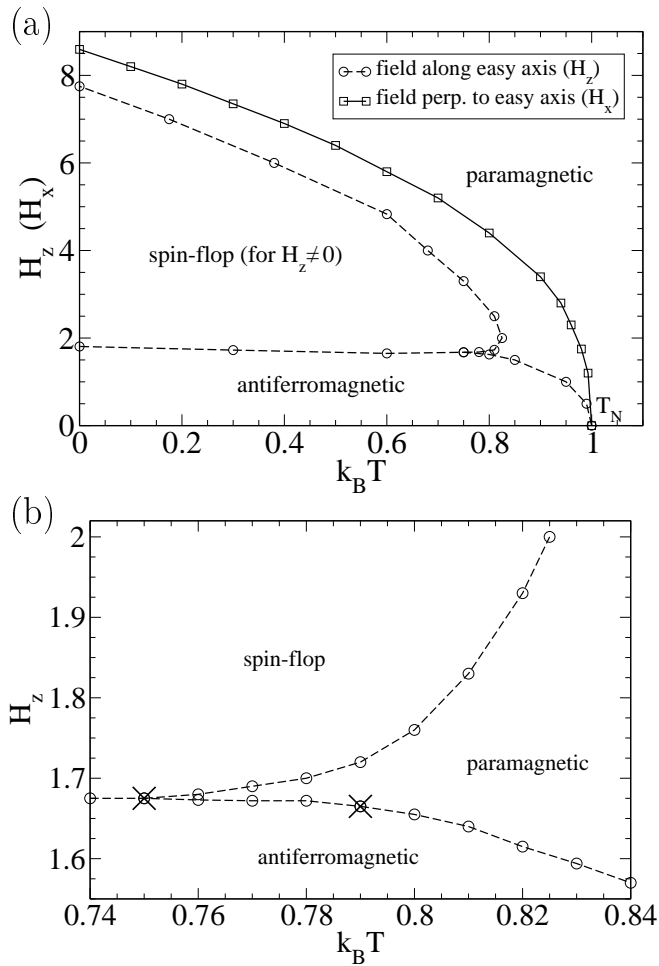}
  \caption{Phase diagram of the model without defects, with fields parallel,
   $H_z$, and perpendicular, $H_x$, to the easy axis.
   (a) Global phase diagram, (b) details in the ($T,H_z$) plane.
   Crosses denote approximate locations of the tricritical point
   and the critical end point (see text).}
  \label{fig_phdiag}
\end{figure}

We analysed the anisotropic Heisenberg model
of Matsuda \textit{et al.}\cite{m4} applying external fields
along the easy axis ($H_z$) and perpendicular to it (say, $H_x$),
and varying the temperature, see the Hamiltonian (\ref{Ham}).
The resulting phase diagrams are depicted in Fig.\ \ref{fig_phdiag}. 

In the case of a field $H_x>0$ perpendicular to the easy axis one encounters,
at zero temperature and small fields, $H_x < H_x^{\mathrm{pm}}$,
an antiferromagnetic ground state with the non-zero, field-dependent
$z$--component of the spins in each chain pointing in the same direction
and alternating sign from chain to chain, $S_{l,m}^z=-S_{l,m+1}^z$.
At zero temperature and larger fields, $H_x > H_x^{\mathrm{pm}}$, the magnetic
field term dominates, and the spins are aligned  along the direction
of the field, $S_{l,m}^x$=1.
The critical field $H_x^{\mathrm{pm}}$ is readily calculated.
Inserting the values of the intrachain, $J_{c1}$ and $J_{c2}$,
and interchain coupling constants, $J_{ac1}$ and $J_{ac2}$, as well
as the spin anisotropy, $D$, as stated in the preceeding section,
one gets $H_x^{\mathrm{pm}}=8.594..$ meV. Certainly, this is an
artificial unit, which had to be transcribed into the standard unit Tesla,
taking into account the $g$-factor and the actual spin value,
when comparing results for non-vanishing fields quantitatively
to experimental findings. However, in the context of our analysis
the artificial unit will be sufficient.
In the following, the unit "meV" will be suppressed in all expressions
for the energy, $k_B$ times temperature ($k_BT$), and the magnetic field.

\begin{figure}[h]
  \includegraphics{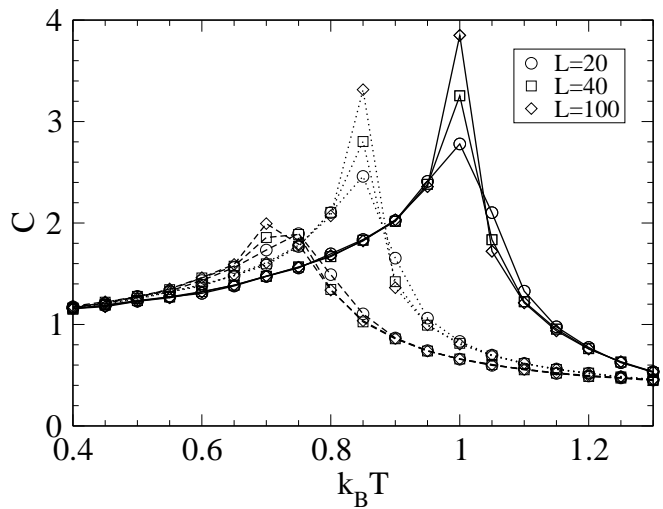}
  \caption{Specific heat vs.\ temperature at fixed fields $H_x=H_z=0$
   (solid lines), $H_x=4.0$ (dotted lines), and $H_z=4.0$ (dashed lines),
   for systems, without defects, of various sizes $L$.}
  \label{fig_spheat}
\end{figure}

\begin{figure}[h]
  \includegraphics{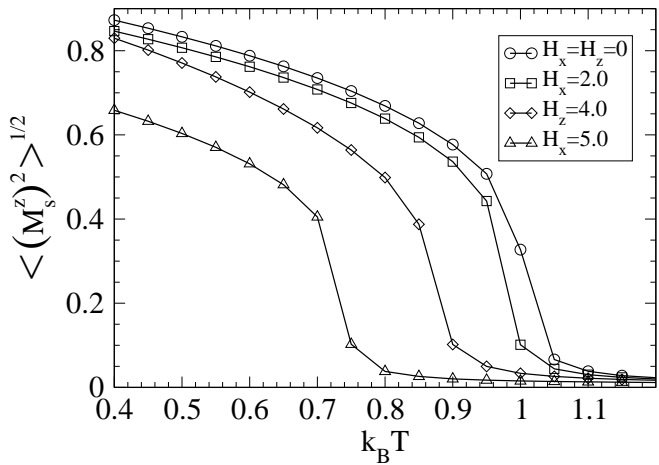}
  \caption{Staggered magnetization vs.\ temperature for various fields
   parallel ($H_z$) and perpendicular ($H_x$) to the easy axis, 
   simulating systems, without defects, of size $L=100$.}
  \label{fig_Mzstagg}
\end{figure}

At non-zero temperatures, a critical line arises from
$(T=0,H_x=H_x^{\mathrm{pm}})$ ending at $(T=T_N,H_x=0)$,
see Fig.\ \ref{fig_phdiag}a.
The transition separates the ordered antiferromagnetic phase
with a non-zero staggered magnetization $<|M_s^z|>$,
see Eq.\ (\ref{Mzstagg}), from the disordered (paramagnetic) phase,
where $<|M_s^z|>=0$. The phase transition is expected to be continuous
and of Ising type, i.e.\ with the well-known critical exponents
of the two-dimensional Ising model. The critical line has been obtained
by fixing either the temperature and varying the field
or by fixing the field and varying the temperature.
Then standard finite-size analyses on the peak positions of the specific heat,
$T_m^C(L)$, were done. Indeed, these positions approach,
for sufficiently large system sizes, the critical temperature $T_c$
of the infinite system like $T_c-T_m^C(L) \propto 1/L$, which is consistent
with the transition belonging to the Ising universality class.
For illustrative purposes, some raw data on the specific heat are shown
in Fig.\ \ref{fig_spheat}.
The thermal behavior of the staggered magnetization is shown,
for a few selected examples, in Fig.\ \ref{fig_Mzstagg}.
 
In the case of an external field $H_z>0$ along the easy axis one obtains
a more complex and more interesting phase diagram, see Fig.\ \ref{fig_phdiag}.
In the ground state ($T=0$), one has to distinguish two critical fields,
$H_z^{\mathrm{sf}}$ and $H_z^{\mathrm{pm}}$. For $H_z < H_z^{\mathrm{sf}}$,
the antiferromagnetic structure, as described above, has the lowest energy.
At larger fields, $H_z^{\mathrm{sf}} < H_z < H_z^{\mathrm{pm}}$,
the spin-flop state is stable. There the $z$--component of the spins
in all chains acquires the same, field-dependent value $S^z(H_z)>0$.
The planar, $xy$--components of the spins are aligned parallel to each other
in each chain, pointing in an arbitrary direction due to the rotational
invariance of the interactions in the $xy$--plane, see Eq.\ (\ref{Ham}).
The $xy$--components of spins in neighboring chains point in the opposite
direction because of the antiferromagnetic interchain couplings.
At $H>H_z^{\mathrm{pm}}$, one has a ferromagnetic ordering with $S^z=1$. 

For the set of couplings obtained from the spin-wave analysis,
the critical fields are $H_z^{\mathrm{sf}}=1.808..$
and $H_z^{\mathrm{pm}}=7.75$.
At $H_z=H_z^{\mathrm{sf}}$, the $z$--component takes the value
$S^z(H_z^{\mathrm{sf}})=0.233..$, corresponding to an angle of $76.5..$ degrees
formed by the $z$ axis and the orientation of the spins.

The complete phase diagram in the $(T,H_z)$ plane consists
of the antiferromagnetic, the spin-flop and the paramagnetic
or disordered states, see Figs.\ \ref{fig_phdiag}a and \ref{fig_phdiag}b.
The antiferromagnetic phase exhibits long-range order
with the staggered magnetization $M_s^z$ as order parameter.
The spin-flop phase has been argued to be of Kosterlitz-Thouless
character,\cite{fn,bl,cp,sch} where transverse spin correlations,
i.e.\ $<S_{lm}^xS_{l'm'}^x + S_{lm}^yS_{l'm'}^y>$, decay algebraically
with distance $\sqrt{(l-l')^2 + (m-m')^2}$.
Accordingly, the transverse sublattice magnetization, see Eq.\ (\ref{Mxy}),
being the order parameter of the spin-flop phase in three dimensions,
is expected to behave for $T>0$ and sufficiently large systems as
\begin{equation}
  \label{g}
  < M_{xy}^2> \propto L^{-g}
\end{equation}
with $g$ approaching 1/4 at the transition from the spin-flop
to the paramagnetic phase,\cite{kt} and $g=2$ in the paramagnetic phase.
Thence, the order parameter vanishes in the Kosterlitz-Thouless phase
as $L \longrightarrow \infty$ at all temperatures $T>0$. Of course,
in the disordered phase spin correlations decay exponentially with distance.

While the existence of these phases for weakly an\-isotropic Heisenberg
antiferromagnets in two dimensions is undisputed, basic aspects
of the topology of the phase diagram and especially the transitions
between the antiferromagnetic phase and the spin-flop as well as
the paramagnetic phases have been discussed controversially,\cite{bl,cp,sch}
and they may, indeed, depend on details of the model.

We determined the boundary line of the antiferromagnetic phase
by monitoring especially the specific heat $C$,
the (square of the) staggered magnetization,
$<|M_s^z|>$ and $<(M_s^z)^2>$, the staggered susceptibility, $\chi_s^z$,
and the Binder cumulant, $U_L$, Eq.\ (\ref{BindCum}). A few raw data
for the specific heat and the staggered magnetization are included
in Figs.\ \ref{fig_spheat} and \ref{fig_Mzstagg}.

The transition from the antiferromagnetic to the disordered phase
at low fields and high temperatures  is continuous and of Ising type.
Its location, as displayed in Figs.\ \ref{fig_phdiag}a and \ref{fig_phdiag}b,
follows from finite-size analyses of the various physical quantities.
The data are consistent with a logarithmic divergence of the specific heat
as well as with the well-known Ising values for the critical exponents
of the order parameter, $\beta=1/8$, and of the staggered susceptibility,
$\gamma=7/4$. 

More interestingly, the transition from the antiferromagnetic
to the paramagnetic phase eventually becomes first order when increasing
the field and lowering the transition temperature, with a tricritical point
at $k_BT_{\mathrm{tr}}=0.79\pm0.015$ and $H_z^{\mathrm{tr}}=1.665\pm0.01$.
The boundary line of the antiferromagnetic phase remains to be first order
at lower temperatures when separating the antiferromagnetic
and the spin-flop phase. The Kosterlitz-Thouless line separating
the spin-flop phase from the paramagnetic (disordered) state hits
the boundary of the antiferromagnetic phase in a critical end point
at $k_BT_{\mathrm{cep}}=0.75\pm0.015$ and $H_z^{\mathrm{cep}}=1.675\pm0.01$,
see Fig.\ \ref{fig_phdiag}b. Note that the phase diagram has qualitatively
the same topology as the one suggested for the spin--1/2 quantum version
of the standard nearest-neighbor antiferromagnet with exchange anisotropy
in two dimensions,\cite{sch} in agreement with the classical version
of that model as well.\cite{hls} Therefore, we conclude that quantum effects
are of minor importance for the main features of the phase diagram. 

\begin{figure}[h]
  \includegraphics{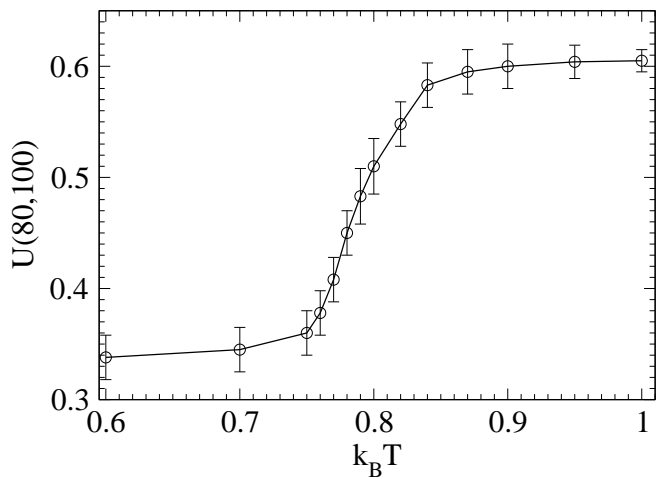}
  \caption{Binder cumulant $U(80,100)$, close to the boundary line 
   of the antiferromagnetic phase, as a function of temperature.}
  \label{fig_BindCum}
\end{figure}

The tricritical point may be located by studying the Binder cumulant.
In the thermodynamic limit the value of the cumulant at the transition point,
$U_{L=\infty}$, is known to depend on the type and universality class
of the transition. In simulations, $U_{L=\infty}$ can be estimated
from systematic finite-size extrapolations of the intersection values
of the Binder cumulant for different system sizes $L_1$ and $L_2$,
$U_{L_1}=U_{L_2}=U(L_1,L_2)$.\cite{bin} In Fig.\ \ref{fig_BindCum},
we depict results for $U(80,100)$, obtained usually at fixed temperature
and varying the field in the vicinity of the boundary
of the antiferromagnetic phase. Obviously, $U(80,100)$ is nearly constant
at high temperatures, $U\approx0.6$, with a fairly rapid change
around $k_BT\approx0.80$. This finding and further finite-size analyses
on $U(L_1,L_2)$ for other system sizes allow us, indeed, to approximately
locate the tricritical point which separates the transition of Ising type,
where $U_{L=\infty}=0.6106..$,\cite{bloe,nico}
and the transition of first order.
Note that the value of $U_{L=\infty}$ may be slightly affected
due to the interactions $J_{c1}$, $J_{c2}$ and $J_{ac2}$.
If only the predominant coupling $J_{ac1}$ were non-zero, the model is easily
seen to be equivalent to a nearest-neighbor Heisenberg antiferromagnet
on a square lattice (cf.\ Fig.\ \ref{fig_Matmodel}).
Clearly, the Hamiltonian then respects the full symmetry of the lattice.
Any of the couplings $J_{c1}$, $J_{c2}$ and $J_{ac2}$ destroys
this lattice isotropy, leading to a spatially anisotropic system
for which cumulant ratios usually exhibit (small) deviations
from their ``isotropic'' values.\cite{bloe,cd}

\begin{figure}[h]
  \includegraphics{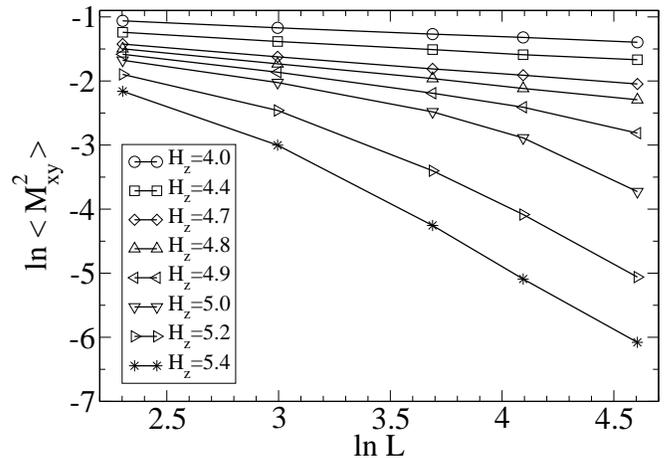}
  \caption{Logarithm of the transverse sublattice magnetization $<M_{xy}^2>$
   versus the logarithm of the system size $L$ at fixed temperature
   $k_BT=0.6$ and for various fields $H_z$, close to the boundary
   between the spin-flop and the disordered phase.}
  \label{fig_Mxy}
\end{figure}

To determine the boundary of the spin-flop phase, we analysed
the size dependence of the transverse sublattice magnetization, $<M_{xy}^2>$.
We apply the criterion that the exponent $g$, see Eq.\ (\ref{g}), is 1/4
at the transition. Typical data are shown in Fig.\ \ref{fig_Mxy},
demonstrating that the magnetization decays much more rapidly with system size
in the paramagnetic phase than in the spin-flop phase. To estimate
the transition point, we determined the local slope
(in a double logarithmic plot),
\begin{equation}
  g_\mathrm{eff}(L) = -\frac{\mathrm{d}\ln<M_{xy}^2>}{\mathrm{d}\ln L},
\end{equation}
from two consecutive system sizes, typically, $L$ and $L+20$.
Indeed, when crossing the phase boundary by fixing the temperature
and lowering the field, $g_\mathrm{eff}(L)$, for large $L$,
tends to jump from 2, characterizing the decay in the disordered phase,
to 1/4 at the transition to the spin-flop phase.
Deeper in the spin-flop phase, $g_\mathrm{eff}$ decreases slightly.

The Kosterlitz-Thouless character of the transition between the spin-flop
and the para\-magnetic phases is also reflected in the thermal behavior
of the specific heat $C$, which displays a non-critical maximum close to,
but not exactly at the transition. Of course, from simulational data
one cannot identify the expected essential singularity of $C$
at the transition.

\section{Effects of defects}\label{sec_defects}
As discussed before,\cite{ls} experiments on La$_5$Ca$_9$Cu$_{24}$O$_{41}$
in a field $H_z$ along the easy axis provide no evidence
for a sharp transition from the antiferromagnetic to the spin-flop phase.
Instead, when fixing the temperature and increasing the field,
the antiferromagnetic phase eventually becomes unstable against
the disordered phase, and spin-flop structures seem to occur at higher fields
only locally as indicated by a quite large, but non-critical maximum in the
susceptibility.\cite{am,rk} The reason for this experimentally
observed behavior is not understood yet. Tentatively, one possible explanation
invokes the holes or defects,\cite{spbk} which may drive the transition
and suppress the spin-flop phase.

In the following, we shall explore this possibility by extending
the classical variant of the anisotropic Heisenberg model
of Matsuda \textit{et al.}\ by adding defects
as described in Sec.\ \ref{sec_model}. Actually, ten percent of the lattice
sites will be occupied by these spinless, mobile defects, in accordance
with the experiments.\cite{nu}
We neglect the quantum nature of the holes and do not, e.g.,
include a kinetic energy or ``hopping'' term in the Hamiltonian
as one would normally expect in the case of a doped quantum antiferromagnet.
Of course, quantum effects may play an important role
for the phase behavior of the doped model.
E.g., quantum fluctuations lead to a non-zero mobility of the holes
even at $T=0$ where our spinless defects are static
due to the absence of thermal fluctuations.
Nevertheless, our classical description is believed to provide some guidance
to novel effects induced by the holes.

Without external field ($H_z=0$) in the ground state ($T=0$), the defects
will form straight stripes perpendicular to the chains.
Due to the next-nearest neighbor interactions, $J_{c2}$ and $J_{ac2}$,
the stripes are bunched with two consecutive defects in a chain keeping
the minimum distance of two lattice spacings with one spin in between
the two defects. Such a bunching did not occur in the related
Ising description,\cite{spbk} where only nearest-neighbor couplings
were assumed. In any event, the bunching may be suppressed, for example,
by a pinning of the defects or repulsive interactions between the defects.
The spins, in the ground state, are oriented along the $z$ axis
with an antiferromagnetic ordering from chain to chain,
as in the case without defects.

\begin{figure}[h]
  \includegraphics{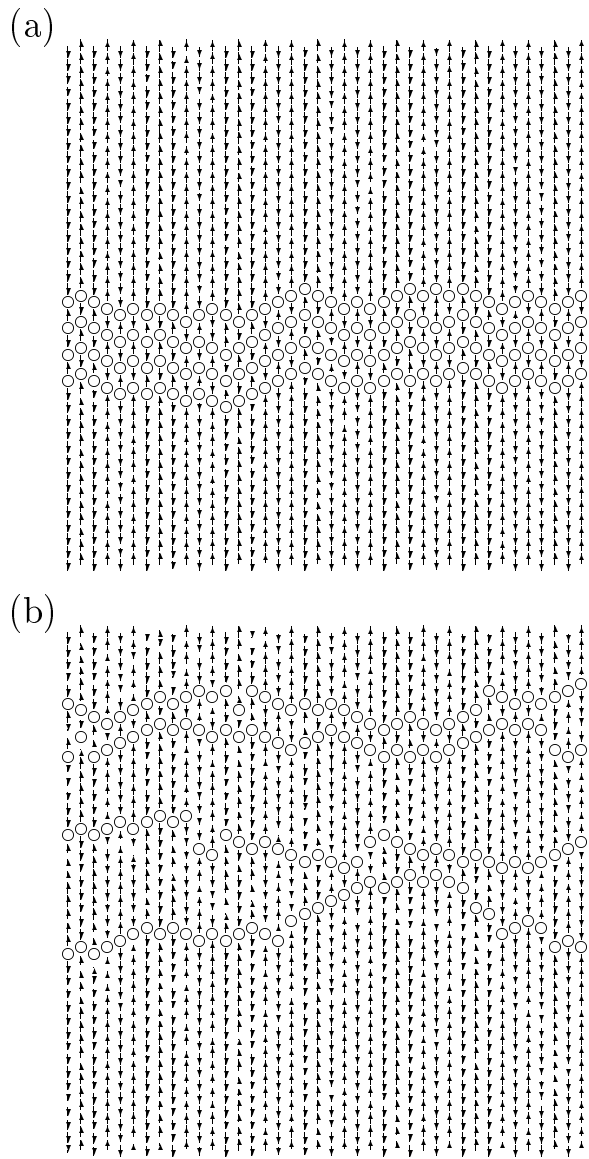}
  \caption{Typical low temperature Monte Carlo equilibrium configurations
   showing (a) at $k_BT=0.3$, the bunching of defect lines,
   and (b) at $k_BT=0.6$, the onset of debunching, both for systems
   of size $L=40$. The open circles denote the mobile defects,
   while the arrows symbolize the $z$--components $S_{lm}^z$ of the spins.}
  \label{fig_equilib_configs}
\end{figure}

At low temperatures, the bunching dominates the typical equilibrium
configurations, as illustrated in Fig.\ \ref{fig_equilib_configs}a.
Upon increasing the temperature, the stripes tend to debunch, thereby gaining
entropy, see Fig.\ \ref{fig_equilib_configs}b. The debunching is reflected
by a steep decrease in the density of defect pairs, i.e.\ consecutive defects
in the same chain separated by merely one spin. The pronounced drop
takes place in a rather narrow range of temperatures
at roughly $k_BT\approx0.55$. However, the debunching seems to be a gradual,
smooth process, without any thermal singularities. 

A phase transition occurs at $k_BT_c\approx0.7$, i.e.\ at a significantly
lower temperature than in the absence of defects. At the transition,
the defect stripes destabilize. As for the Ising model with mobile defects,
the stripe instability may be inferred from the average minimal distance $d_a$
between each defect in chain $m$, at position $(l_d,m)$,
and those in the next chain, at $(l_d',m+1)$, defined by \cite{spbk,hs}
\begin{equation}
  d_a = \sum\limits_{l_d} \langle \min|l_d-l_d'| \rangle / N_d,
\end{equation}
dividing the sum by the number $N_d$ of defects. This quantity increases
rapidly at the transition. The transition is also marked in the simulations
by a pronounced peak in the specific heat and a drastic decrease
in the sublattice magnetization, which is expected to vanish
at $T\ge T_c$ in the thermodynamic limit. 

\begin{figure}[h]
  \includegraphics{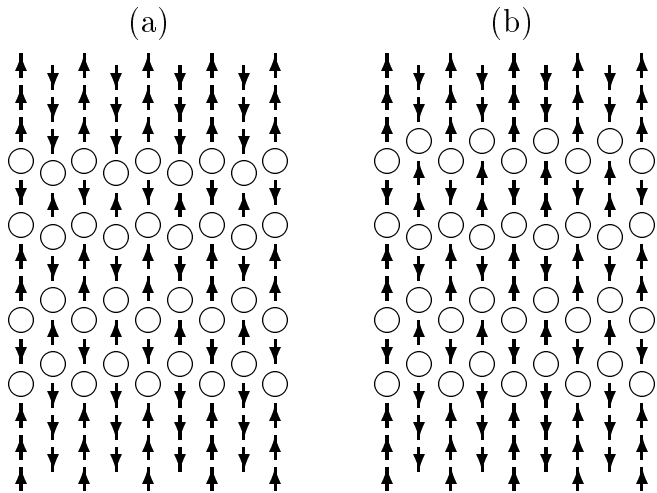}
  \caption{Illustration of the ground state configurations in an external
   field $H_z>0$ for a system with four defects per chain;
   for (a) $H_z^{(1)}<H_z<H_z^{(2)}$, and (b) $H_z^{(2)}<H_z<H^{\mathrm{sf}}$.}
  \label{fig_grndstate}
\end{figure}

Applying an increasing external field $H_z>0$ along the easy axis, the results
of the ground state calculation (for even numbers of at least four defects
per chain) may be summarized as follows.
First, one has to distinguish two fields $H_z^{(1)}$ and $H_z^{(2)}$.
For $0<H_z<H_z^{(1)}$ one keeps the same antiferromagnetic structure
with bunched defect stripes as in the case of vanishing field.
Then, for $H_z^{(1)}<H_z<H_z^{(2)}$, precisely one additional spin
(pointing along the field direction) is inserted between
two consecutive defects in every other chain, see Fig.\ \ref{fig_grndstate}a.
This configuration becomes unstable at $H_z=H_z^{(2)}$,
and now an additional spin pointing in the $z$ direction is inserted
between two consecutive defects in every chain (Fig.\ \ref{fig_grndstate}b).
For simplicity and by analogy to the Ising case,\cite{spbk} we refer
to these two ground state configurations as "zig-zag" structures.
The two fields $H_z^{(1)}$ and $H_z^{(2)}$ are readily found to be given
by $H_z^{(1)}=-J_{ac2}+J_{c2}/2$ and $H_z^{(2)}=2H_z^{(1)}$.
Inserting the values of the interaction constants
(see Sec.\ \ref{sec_model}) one obtains $H_z^{(1)}=0.4305$
and $H_z^{(2)}=0.861$. Note that broader regions of "inserted" spins are,
however, not favored energetically even when increasing the field.
Instead, for larger fields, eventually a spin-flop transition occurs
at $H^{\mathrm{sf}}$, followed by a ferromagnetic structure at higher fields,
as in the case without defects.

\begin{figure}[h]
  \includegraphics{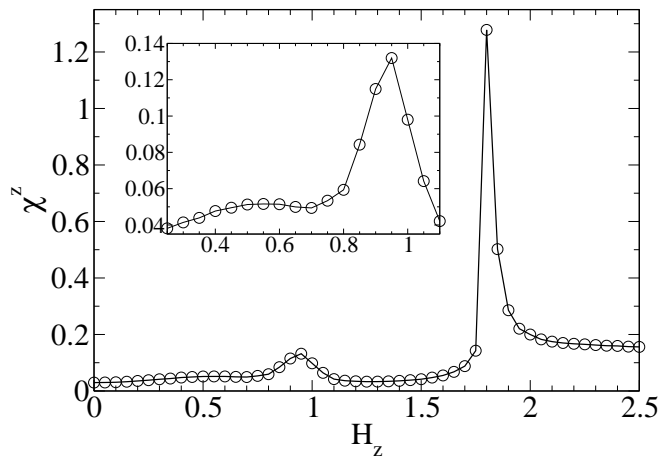}
  \caption{Susceptibility $\chi^z$ as a function of the magnetic field $H_z$
   at fixed temperature $k_BT=0.3$, for a system of size $L=40$.
   The inset shows the existence of a very weak maximum at $H_z\approx0.5$.}
  \label{fig_chiz}
\end{figure}

The zig-zag structures lead to a stepwise increase in the total magnetization,
which gives rise to rather small maxima in the susceptibility $\chi^z$
at low temperatures, as depicted in Fig.\ \ref{fig_chiz}.
One observes a small, non-critical peak at $H_z\approx0.9$,
which is the remnant of the transition at $T=0$ between the two
zig-zag structures depicted in Fig.\ \ref{fig_grndstate}.
Moreover, a very weak, non-critical maximum can be identified
at $H_z\approx0.5$ (see inset). This peak is associated
with the zig-zag structure of Fig.\ \ref{fig_grndstate}a.
At higher fields, a pronounced peak occurs
signaling the transition to the spin-flop phase.

We found, however, no evidence for a phase transition at finite temperatures
associated with the small peaks in $\chi^z$. Instead, upon increasing the
field, straight stripes seem to transform gradually into zig-zag stripes,
as for the Ising model with mobile defects.\cite{mh}
The first peaks already vanishes at about $k_BT\approx0.3$,
and the position of the second small peak shifts to somewhat higher fields
and gets less pronounced as the temperature is increased.
It disappears at about  $k_BT\approx 0.5$, possibly due to the debunching.
Obviously, the occurrence of zig-zag structures cannot be identified
with the phase transition in
La$_5$Ca$_9$Cu$_{24}$O$_{41}$ observed well below
the onset of spin-flop structures.
 
Indeed, the anisotropic Heisenberg model with mobile defects still displays,
at low temperatures, a sharp transition from the antiferromagnetic phase,
with straight or zig-zag defect stripes, to a spin-flop phase, as signaled
by a delta-like peak in the susceptibility $\chi^z$ (see Fig.\ \ref{fig_chiz}).
The topology of the phase diagram in the $(T,H_z)$ plane seems to resemble
that in the absence of defects, see Figs.\ \ref{fig_phdiag}a
and \ref{fig_phdiag}b. Actually, at the triple point (or critical end point)
between the antiferromagnetic, spin-flop and paramagnetic phases,
located roughly at $(k_BT=0.5, H_z=1.7)$, the, presumably, non-critical
debunching line seems to meet as well. However, we did not attempt to map
the phase diagram accurately, because obviously the introduction of defects
does not suffice to reconcile the experimental findings on
La$_5$Ca$_9$Cu$_{24}$O$_{41}$, showing no direct
transition from the antiferromagnetic to the spin-flop phase.
In fact, the possible destruction of the spin-flop phase by the instability
of the defect stripes tends to be hindered by the bunching of the stripes.
Further investigations are desirable, but beyond
the scope of the present study.

\section{Summary}
We have analysed in detail a classical variant of a two-dimensional
Heisenberg antiferromagnet with weak, uniaxial anisotropy
proposed by Matsuda \textit{et al.}\ to reproduce spin-wave dispersions
measured in the magnet La$_5$Ca$_9$Cu$_{24}$O$_{41}$.
In particular, we determined the phase diagrams of the model applying fields
parallel, $H_z$, and perpendicular, $H_x$, to the easy axis
of the spin anisotropy.

In the case of a transverse field $H_x$ (perpendicular to the easy axis),
the transition from the antiferromagnetic phase to the paramagnetic phase
belongs to the Ising universality class.
The phase diagram in the case of a field $H_z$ pointing along the easy axis
consists of the antiferromagnetic, the spin-flop, and the disordered
(paramagnetic) phases. Extensive analyses have been performed to locate
the phase boundaries, partly motivated by conflicting analyses
of related models. Indeed, our analysis, studying especially
the Binder cumulant and the transverse sublattice magnetization,
allows one to locate reasonably well both the tricritical point
on the phase boundary between the antiferromagnetic
and the paramagnetic phases as well as the critical end point
between these two phases and the spin-flop phase. Quantum effects seem to play
no essential role for the topology of the phase diagram,
which is in qualitative disagreement with experimental observations
on La$_5$Ca$_9$Cu$_{24}$O$_{41}$.

We extended the classical variant of the model
of Matsuda \textit{et al.}\ by including spinless mobile defects
mimicking the holes in La$_5$Ca$_9$Cu$_{24}$O$_{41}$,
thereby following previous suggestions on a related Ising model.
In the antiferromagnetic phase, the defects, at low temperatures
and low fields, are found to form stripes as in the corresponding Ising case.
However, due to next-nearest neighbor couplings, the stripes tend to bunch.
The debunching, occurring at higher temperatures, seems to be non-critical,
albeit it takes place in a rather narrow temperature range.
A phase transition at which the antiferromagnetic order is destroyed
is driven by a destruction of the defect stripes loosing their coherency
at the transition.

The model with defects has also been studied in the presence of a field
along the easy axis. There, a spin-flop phase is observed as well,
separated from the antiferromagnetic phase presumably by a transition
of first order. Therefore, we conclude that adding the mobile defects
is not sufficient to reconcile model properties with experimental findings
ruling out a direct transition from the antiferromagnetic
to the spin-flop phase. Perhaps a destruction of the spin-flop phase may occur
when the bunching is suppressed.

However, when interpreting our findings for the model with defects
one should keep in mind that our description of the holes
is a purely classical one.
Quantum fluctuations reduce the clustering tendency of the holes
and may also destroy the bunched structures
that we find from our classical ground state analysis.
Thus the role of quantum effects should certainly
be investigated more carefully when comparing with actual experiments.

In any event, the models display various interesting behavior,
and they may well contribute to arriving at a really satisfying
theoretical description of the intriguing experimental observations
on the La$_5$Ca$_9$Cu$_{24}$O$_{41}$ magnets.
Moreover, the methods used in our study may be helpful in analysing
phase diagrams of other two-dimensional weakly anisotropic
Heisenberg antiferromagnets.

\acknowledgments
It is a pleasure to thank B.\ B\"uchner, M.\ Holtschneider,
and R.\ Klingeler for very useful discussions, information, and help.
An inspiring correspondence with M.\ Matsuda is also much appreciated.
Financial support by the Deutsche Forschungsgemeinschaft
under grant No.\ SE324 is gratefully acknowledged.

\end{document}